\documentclass[preprint]{revtex4}
\usepackage{amsfonts}
\usepackage{amssymb,epsfig}
\usepackage{latexsym}
\usepackage{amsmath}
\usepackage{graphicx}

\begin{document}

\title{Interacting entropy-corrected holographic dark
energy with apparent horizon as an infrared cutoff}
\author{A. Khodam-Mohammadi\footnote{Email:\text{khodam@basu.ac.ir}}~~and~~
M. Malekjani\footnote{Email:\text{malekjani@basu.ac.ir}}}

\affiliation{Department of Physics, Faculty of Science, Bu-Ali Sina
University, Hamedan 65178, Iran}

\begin{abstract}
In this work we consider the entropy-corrected version of
interacting holographic dark energy (HDE), in the non-flat universe
enclosed by apparent horizon. Two corrections of entropy so-called
logarithmic 'LEC' and power-law 'PLEC' in HDE model with apparent
horizon as an IR-cutoff are studied. The ratio of dark matter to
dark energy densities $u$, equation of state parameter $w_D$ and
deceleration parameter $q$ are obtained. We show that the cosmic
coincidence is satisfied for both interacting models. By studying
the effect of interaction in EoS parameter, we see that the phantom
divide may be crossed and also find that the interacting models can
drive an acceleration expansion at the present and future, while in
non-interacting case, this expansion can happen only at the early
time. The graphs of deceleration parameter for interacting models,
show that the present acceleration expansion is preceded by a
sufficiently long period deceleration at past. Moreover, the
thermodynamical interpretation of interaction between LECHDE and
dark matter is described. We obtain a relation between the
interaction term of dark components and thermal fluctuation in a
non-flat universe, bounded by the apparent horizon. In limiting
case, for ordinary HDE, the relation of interaction term versus
thermal fluctuation is also calculated.
\end{abstract}
\maketitle
\section{Introduction}
The dark energy scenario has attracted a great deal of attention in
the last decade. Many cosmological observations reveal that our
universe evolves under an acceleration expansion \cite{Perl}. This
expansion may be driven by an unknown energy component with negative
pressure, so called, dark energy (DE), which fills $\sim70$ percent
of energy content of our universe with an effective equation of
state (EoS) parameter $-1.48<w_{eff}<-0.72$ \cite{Hannestad}.
Despite of many efforts in this subject, the nature of DE is the
most mysterious problem in modern cosmology. The first and simplest
candidate of dark energy is $\Lambda$CDM model, in which
$w_{\Lambda}=-1$ is constant. Although this model is consistent very
well with all observations, it faces with with the fine tuning and
cosmic coincidence problem. After this, the dynamical DE models have
been proposed to solve the DE problems. Among many dynamical models
of DE, in which $w_D$ is not constant, the entropy-corrected dark
energy models based on quantum field theory and gravitation have
been widely extended by many authors in recent years
\cite{Pavon5,E-C}. The motivation of these corrections has been
based on black hole physics, where some gravitational fluctuations
and field anomalies can affect the entropy-area law of black holes.
The logarithmic and power-law corrections of entropy are two
procedures in dealing with this fluctuations. First correction has
been given by logarithmic fluctuations at the spacetime, in the
context of loop quantum gravity (LQG) \cite{LQG}. The entropy-area
relationship leads to the curvature correction in the
Einstein-Hilbert action and vice versa \cite{zhu,Ody3}. In this case
the corrected entropy is given by \cite{modak}
\begin{equation}
S_{\mathrm{BH}}=\frac{A}{4G}+\tilde{\gamma}\ln{\frac{A}{4G}}+\tilde{\beta},
\label{MEAR}
\end{equation}
where $\tilde{\gamma}$ and $\tilde{\beta}$ are dimensionless
constants of order unity. By considering the entropy correction, the
energy density of logarithmic entropy-corrected holographic dark
energy (LECHDE) can be given as \cite{wei1}
\begin{equation}
\rho _{D}=3n^2M_{p}^{2}L^{-2}+\gamma L^{-4}\ln
(M_{p}^{2}L^{2})+\beta L^{-4}.\label{rhoS}
\end{equation}
Three parameters $n,~\beta$ and $\gamma$ are parameters of model and
$M_P$ is the reduced Planck mass. The correction terms (two last
terms of (\ref{rhoS})) are effective only at the early stage of the
universe and they will be vanished when the universe becomes large,
in which $\rho _{D}^{EC}\rightarrow \rho _{D}^{O}$, where $\rho
_{\Lambda}^{O}=3n^2M_{p}^{2}L^{-2}$ is the dark energy density of
ordinary HDE model (more discussion of HDE model is referred to
\cite{HDE}). In this model, the IR-cutoff '$L$' plays an essential
role. If $L$ is chosen as particle horizon, the HDE can not make an
acceleration expansion \cite{hsu}, while for future event horizon,
Hubble scale '$L=H^{-1}$', and apparent horizon (AH) as an
IR-cutoff, the HDE can simultaneously drive accelerated expansion
and solve the coincidence problem \cite{Pavon1,28sheykhi,Sheykhi1}.
More recently, a model of interacting HDE (i.e. a non gravitational
interaction between DE and dark matter (DM)) at Ricci scale, in
which $L=(\dot{H}+2H^2)^{-1/2}$ has been proposed. The authors
performed a detailed discussion on the cosmic coincidence problem,
age problem and obtained some observational constraints on their's
model \cite{Pavon2}.

The second class of ECHDE, power-law correction of entropy (PLEC),
is appeared in dealing with the entanglement of quantum fields in
and out of the horizon \cite{Das2}. In this model, the
corrected-entropy is given by \cite{Pavon5}
\begin{equation}
S=\frac{A}{4G}[1-K_{\alpha}A^{1-\alpha/2}],\label{entpl}
\end{equation}
where $\alpha$ is a dimensionless positive constant and
\begin{equation}
K_{\alpha}=\frac{\alpha}{4-\alpha}(4\pi r_{c}^2)^{\alpha/2-1}.
\label{kalpha}
\end{equation}
Here $r_{c}$ is the crossover scale. More detail is referred to
\cite{Pavon5,Das2,Das3}. It is worthwhile to mention that in the
most acceptable range of $4>\alpha>2$ \cite{Pavon5,Das2}, the
correction term (i.e. the second term of (\ref{entpl})), is
effective only at small $A$'s and it falls off rapidly at large
values of $A$. Therefore, for large horizon area, the ordinary
entropy-area law (first term of (\ref{entp1})) is recovered. However
the thermodynamical considerations predict that the case
$\alpha\leq2$ may be acceptable, but as we will show in Sec.
\ref{PLECHDE}, this range should be removed by cosmic coincidence
consideration. Due to entropy corrections to the Bekenstein-Hawking
entropy ($S_{BH}$), the Friedmann equation should be modified
\cite{Pavon5}. In comparison with ordinary Friedman equation, the
energy density of PLECHDE, has been given by \cite{Sheykhi2}
\begin{equation}
\rho_{D}=3n^2M_{p}^{2}L^{-2}-\delta
M_{p}^{2}L^{-\alpha},\label{endpl}
\end{equation}
where $\delta$ and $\alpha$ are the parameters of PLECHDE model. We
must mention that the ordinary HDE is recovered for $\delta=0$ or
$\alpha=2$.

In historical point of view, laws of black hole thermodynamics have
made some relations between thermodynamics and a self gravitating
system bounded by a horizon. In this theory, some thermodynamical
quantities such as entropy and temperature are purely geometrical
quantities which have been obtained from area and surface gravity of
horizon, respectively. In the Friedmann-Robertson-Walker (FRW)
universe, with horizons, like future event horizon in black hole
physics, by studying the thermodynamical quantities and generalized
second law (GSL)\cite{wang}, one can choose the best DE model or the
best horizon. For example, it has been shown that in a non-flat FRW
universe, enclosed by apparent horizon, the GSL is governed
irrespective of any DE model \cite{Sheykhi1}. The investigation of
GSL for LECHDE and PLECHDE models has been performed in
\cite{Pavon5}.

Recently, the HDE and agegraphic/new-agegraphic DE models have been
extended regarding the entropy corrections (LECHDE, PLECHDE,
PLECNADE) and a thermodynamical description of the LECHDE model has
been studied \cite{E-C,wei1,Sheykhi2,khodam3}. Also at Ref.
\cite{Sheykhi1}, thermodynamics interpretation of interacting
holographic dark energy with AH-IR-cutoff, enclosed by apparent
horizon, was studied. These papers give us a strong motivation to
study the LECHDE and PLECHDE models with AH-IR-cutoff in a non-flat
universe, enclosed by apparent horizon, which is a generalization of
earlier works of Sheykhi et.al. \cite{Sheykhi1,Sheykhi2}. It should
be mentioned that, the motivation of a closed universe has been also
shown in a suite CMB experiments \cite{Sie} and of the cubic
correction to the luminosity-distance of supernova measurements
\cite{Caldwell}.

The outline of our paper is as follows: In Sec. \ref{ECHDE}, the
interacting LECHDE model with AH-IR-cutoff is studied and the
evolution of dark energy, deceleration parameter and EoS parameter
are calculated. Also these calculations are performed for PLECHDE
model with AH-IR-cutoff in Sec. \ref{PLECHDE}. In Sec. \ref
{nonint}, the thermodynamical quantities such as entropy and Hawking
temperature of apparent horizon are obtained only for LECHDE model
and then the interaction term due to thermal fluctuation is obtained
in Sec. \ref{int}. We finish Our paper with some concluding remarks.

\section{Interacting ``LECHDE" model with AH-IR-cutoff}\label{ECHDE}

The line element of a homogenous and isotropic FRW universe is given
by
\begin{equation}
ds^{2}=h_{ab}dx^{a}dx^{b}+\widetilde{r}^{2}(d\theta ^{2}+\sin ^{2}\theta
d\phi ^{2}),
\end{equation}%
where $\widetilde{r}=a(t)r,$ two non-angular metric
$(x^{0},x^{1})=(t,r)$
and two dimensional metric is $h_{ab}=diag[-1,a^{2}/(1-Kr^{2})]$. Here $%
K=1,0,-1 $ is the curvature parameter corresponding to a closed, flat and
open universe, respectively. The dynamical apparent horizon, a marginally
trapped surface with vanishing expansion is $\widetilde{r}%
_{A}=(H^{2}+K/a^{2})^{-1/2}$ which has been calculated by the relation $%
h^{ab}\partial _{a}\widetilde{r}\partial _{b}\widetilde{r}=0$
\cite{Cai1}. This relation implies that the vector $\nabla
\widetilde{r}$ is null on the apparent horizon surface. The apparent
horizon may be considered as a causal horizon for a dynamical
spacetime. Thus one can associate a gravitational entropy and
surface gravity to it \cite{Bak}.

From Eq. (\ref{rhoS}), the energy density of LECHDE with apparent
horizon, $\widetilde{r}_{A}$, as an IR-cutoff can be written as%
\begin{equation}
\rho _{D}=3n^{2}M_{P}^{2}\widetilde{r}_{A}^{-2}+\gamma \widetilde{r}%
_{A}^{-4}\ln (M_{P}^{2}\widetilde{r}_{A}^{2})+\beta
\widetilde{r}_{A}^{-4}. \label{ED}
\end{equation}
The first Friedmann equation is%
\begin{equation}
\frac{1}{\widetilde{r}_{A}^{2}}=H^{2}+\frac{K}{a^{2}}=\frac{1}{3M_{P}^{2}}%
(\rho _{m}+\rho _{D}),  \label{FReq}
\end{equation}%
where $H=\dot{a}/a$ is the Hubble parameter. In a FRW universe, the
total energy density $\rho =\rho _{D}+\rho _{m}$ is satisfied in a
conservation equation as:
\begin{equation}
\dot{\rho}+3H(1+w)\rho =0
\end{equation}%
where $w=p/\rho $ is the EoS parameter. Due to non gravitational
interaction between dark energy and pressureless cold dark matter
(CDM) with subscript '$m$', two energy densities $\rho _{D}$ and
$\rho _{m}$ are not conserved separately and the conservation
equation can be written as
\begin{eqnarray}
\dot{\rho}_{D}+3H(1+w_{D})\rho _{D} &=&-Q,  \label{CE1} \\
\dot{\rho}_{m}+3H\rho _{m} &=&Q.  \label{CE2}
\end{eqnarray}%
Here $Q$ is the interaction term which has been usually considered
in
three forms as \cite{wei3}%
\begin{equation}
Q=\Gamma \rho _{D}=\left\{
\begin{array}{l}
3Hb^{2}\rho _{D} \\
3Hb^{2}\rho _{m} \\
3Hb^{2}(\rho _{m}+\rho _{D})%
\end{array}%
\right\} .  \label{INTQ}
\end{equation}%
In this equation, $b^{2}$ is coupling constant. Although a
theoretical interpretation of this interaction has not been
performed yet, as we see from Eqs. (\ref{CE1}, \ref{CE2}), the
interaction term $Q$ should be as a function of $H$ multiplied to
energy density. Therefore in Eq. (\ref{INTQ}), the simplest form of
$Q$ is considered with a coupling constant $b$. This term indicates
the decay rate of DE to CDM as similar as standard $\Lambda $CDM
model where vacuum fluctuations can decay into matter. In many
models the interaction term is necessary in order to solving the
coincidence problem. It has been shown that this interaction can
influence the perturbation dynamics, cosmic microwave background
(CMB) spectrum and structure formation \cite{jamil27}.

\begin{center}
\begin{figure}[]
\includegraphics[width=9cm]{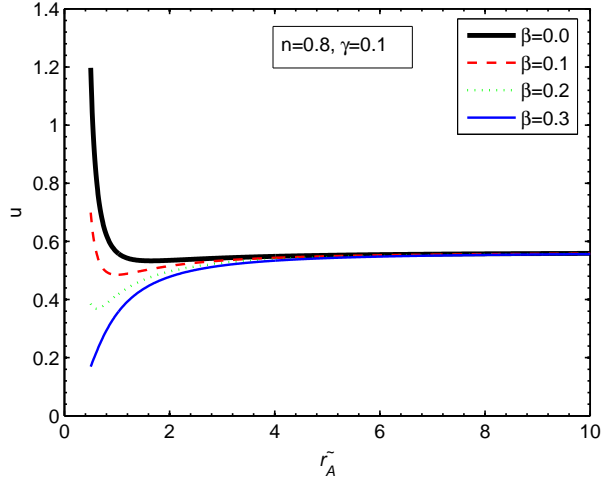}
\caption{The evolution of $u$ in versus $\widetilde{r}_{A}$ in
LECHDE model. The asymptotic value is $u=0.56$.} \label{fig1}
\end{figure}
\end{center}

Differentiating Eq. (\ref{ED}) with respect to cosmic time and using
the
differentiation of apparent horizon with respect to cosmic time, we have%
\begin{equation}
\dot{-\widetilde{r}_{A}}\widetilde{r}_{A}^{-3}=H(\dot{H}-\frac{K}{a^{2}})=%
\frac{1}{6M_{P}^{2}}(\dot{\rho}_{D}+\dot{\rho}_{m}),  \label{eq10}
\end{equation}%
where from Eqs. (\ref{CE1}, \ref{CE2}) we obtain%
\begin{equation}
\dot{\widetilde{r}_{A}}=\frac{H}{2M_{P}^{2}}\widetilde{r}_{A}^{3}\rho
_{D}(1+u+w_{D}), \label{rd}
\end{equation}
\begin{equation}
\dot{\rho}_{D}=-\frac{H\rho _{D}\widetilde{r}_{A}^{2}}{M_{P}^{2}}%
(1+u+w_{D})[2\rho _{D}-\gamma \widetilde{r}_{A}^{-4}-3n^{2}M_{P}^{2}%
\widetilde{r}_{A}^{-2}].  \label{dED}
\end{equation}%
Here $u=\rho _{m}/\rho _{D}$ is the ratio of energy densities. Also
from Eq. (\ref{FReq}), we find that $3M_{P}^{2}\widetilde{r}
_{A}^{-2}=(1+u)\rho _{D}$ where $u$ is governed by
\begin{equation}
u=\frac{3M_{P}^{2}}{3n^{2}M_{P}^{2}+\gamma \widetilde{r}_{A}^{-2}\ln
(M_{P}^{2}\widetilde{r}_{A}^{2})+\beta \widetilde{r}_{A}^{-2}}-1.
\label{eq11}
\end{equation}%
From Eq. (\ref{eq11}), we see that at sufficient large $\widetilde{r}_{A}$, where $%
\rho _{D}\approx 3n^{2}M_{P}^{2}\widetilde{r}_{A}^{-2},$ the ratio
of energy densities will tend to a constant value $u\rightarrow
1/n^{2}-1$. Also at present time, $u$ varies slowly up to reach a
constant value, $u=1/n^{2}-1$. In Fig. \ref{fig1}, the function $u$
is plotted in versus $\widetilde{r}_{A}$ for fixed $\gamma ,~n$ and
various  $\beta$  in the Planck mass unit in which
$M_{P}=1/\sqrt{8\pi G}=1$. From this figure, we conclude that the
coincidence problem gets alleviated since for some values of model
parameters, we get $u\sim\mathcal{O}(1)$ for wide range of
$\widetilde{r}_{A}$ (including the present time), and it is growing
so that finally reaches to a fixed value of order unity.
The deceleration parameter $q=-1-\dot{H}/H^{2}$ may be obtained by
using the Friedmann equation and continuity equation as follows
\cite{28sheykhi,Sheykhi1}
\begin{equation}
q=-(1+\Omega _{K})+\frac{3}{2}\Omega _{D}(1+u+w_{D}),  \label{dec1}
\end{equation}%
where $\Omega _{K}=K/(a^{2}H^{2})$, $\Omega _{D}=\rho
_{D}/(3M_{P}^{2}H^{2})$ and $\Omega _{m}=\rho
_{m}/(3M_{P}^{2}H^{2})$ are the energy density parameters. From
these dimensionless parameters, the first Friedmann equation can be
rewritten as: $1+\Omega _{K}=\Omega _{D}+\Omega _{m}$. Using the
third form of interacting term, in which $\Gamma/3H=b^2(1+u)$ and
combining Eq. (\ref{dED}) with (\ref{CE1}), the EoS parameter
$w_{D}$ is given by
\begin{equation}
w_{D}=-1-\frac{u(2\rho
_{D}-3n^{2}M_{P}^{2}\widetilde{r}_{A}^{-2}-\gamma
\widetilde{r}_{A}^{-4})-b^2(1+u)^2\rho _{D}}{(1-u)\rho
_{D}-3n^{2}M_{P}^{2}\widetilde{r}_{A}^{-2}-\gamma
\widetilde{r}_{A}^{-4}}. \label{EoS}
\end{equation}%
From this equation and Eq. (\ref{eq11}), we find
\begin{eqnarray}
&&\widetilde{r}^{\prime}_{A}=\frac{3M_{P}^2\widetilde{r}_{A}}{2}\Big{[}3n^2M_{P}^2\widetilde{r}_{A}^2+\gamma
\ln(M_{P}^2\widetilde{r}_{A}^2)\notag \\
&&+\beta+3M_{P}^2\widetilde{r}_{A}^2(b^2-1)\Big{]}/ \Big{[}3M_{P}^2\widetilde{r}_{A}^2(n^2-1)\notag \\
&&+2\gamma \ln(M_{P}^2\widetilde{r}_{A}^2)+2\beta-\gamma\Big{]},
\label{rp}
\end{eqnarray}
where ``prime" denotes the differentiation with respect to
$x=\ln~a=-\ln(1+z)$ in which $Hd/dx=d/dt$.

On the other hand, by using Eqs. (\ref{FReq}) and (\ref{INTQ}), the
evolution of dark energy density can be rewritten as
\begin{equation}
\rho^{\prime}_D=-3\rho_D\left[1+w_D+b^2(1+u)\right],  \label{ed2}
\end{equation}
and then the evolution of $\Omega_D$ is calculated as:
\begin{equation}
\Omega^{\prime}_D=-3\Omega_D\left[(1+w_D)(1-\Omega_D)+b^2(1+u)-\Omega_D
u+\frac{2}{3}\Omega_K\right]. \label{ed3}
\end{equation}%
Using Eq. (\ref{dec1}), the deceleration parameter is given by
\begin{equation}
q=-(1+\Omega _{K})-\frac{3}{2}\frac{\Omega _{D}(1+u)[u-b^2(1+u)]\rho
_{D}}{(1-u)\rho _{D}-3n^{2}M_{P}^{2}\widetilde{r}_{A}^{-2}-\gamma
\widetilde{r}_{A}^{-4}}.  \label{dec2}
\end{equation}
It is worthwhile to mention that $\Omega_K$ and $\Omega_D$ is
related by
\begin{equation}
\frac{\Omega_K}{\Omega_m}=a\frac{\Omega_{K_0}}{\Omega_{m_0}}~~\therefore~~
\Omega_K=\frac{e^{x}\Gamma(1-\Omega_D)}{1-e^{x}\Gamma} \label{Ok},
\end{equation}
where $\Gamma=\Omega_{K_0}/\Omega_{m_0} $ is a constant value, which
from the recent data, is given by $\Gamma\approx 0.04$. Here the
subscript '0', is used for the present time.

\begin{center}
\begin{figure}[]
\includegraphics[width=9cm]{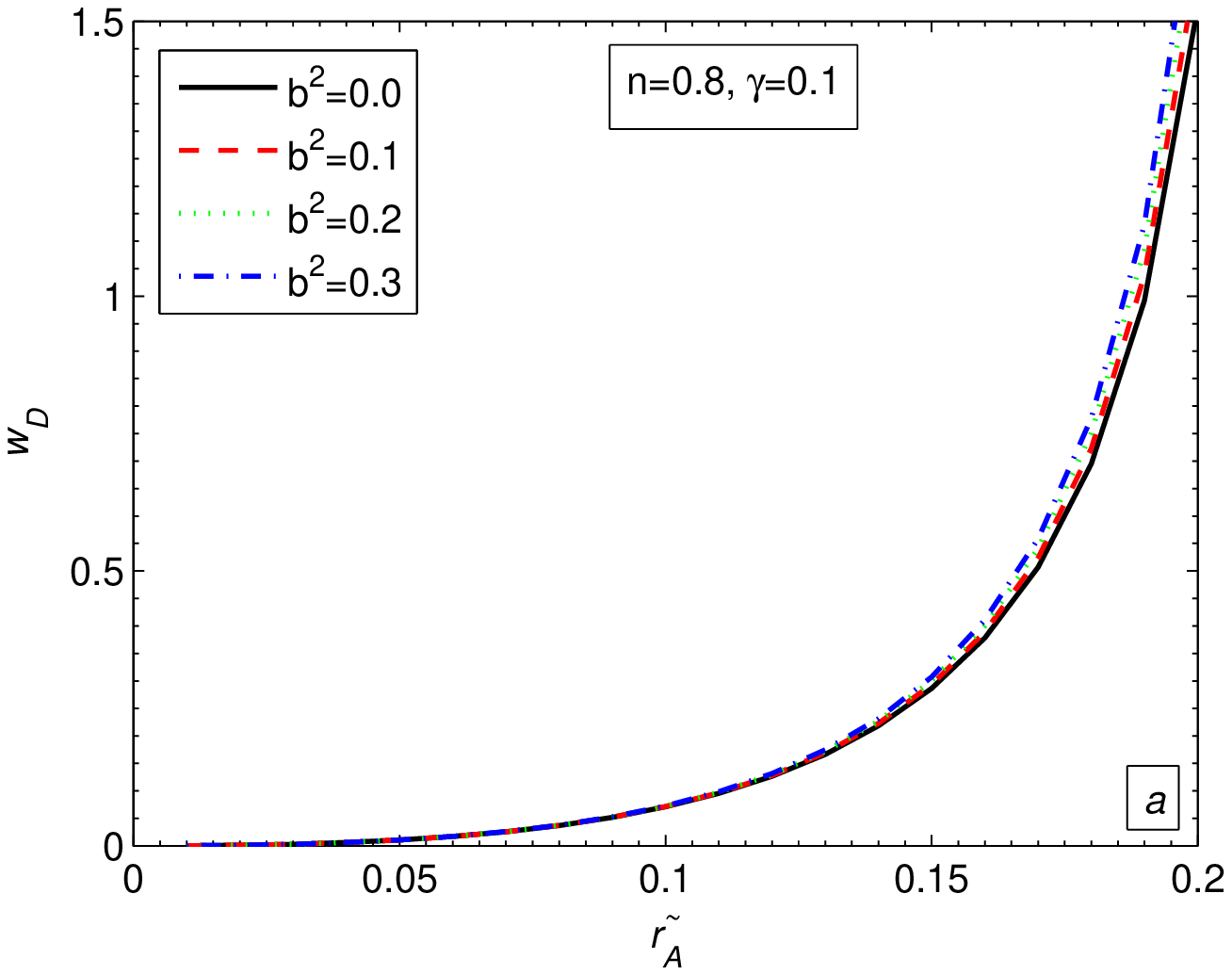}
\includegraphics[width=9cm]{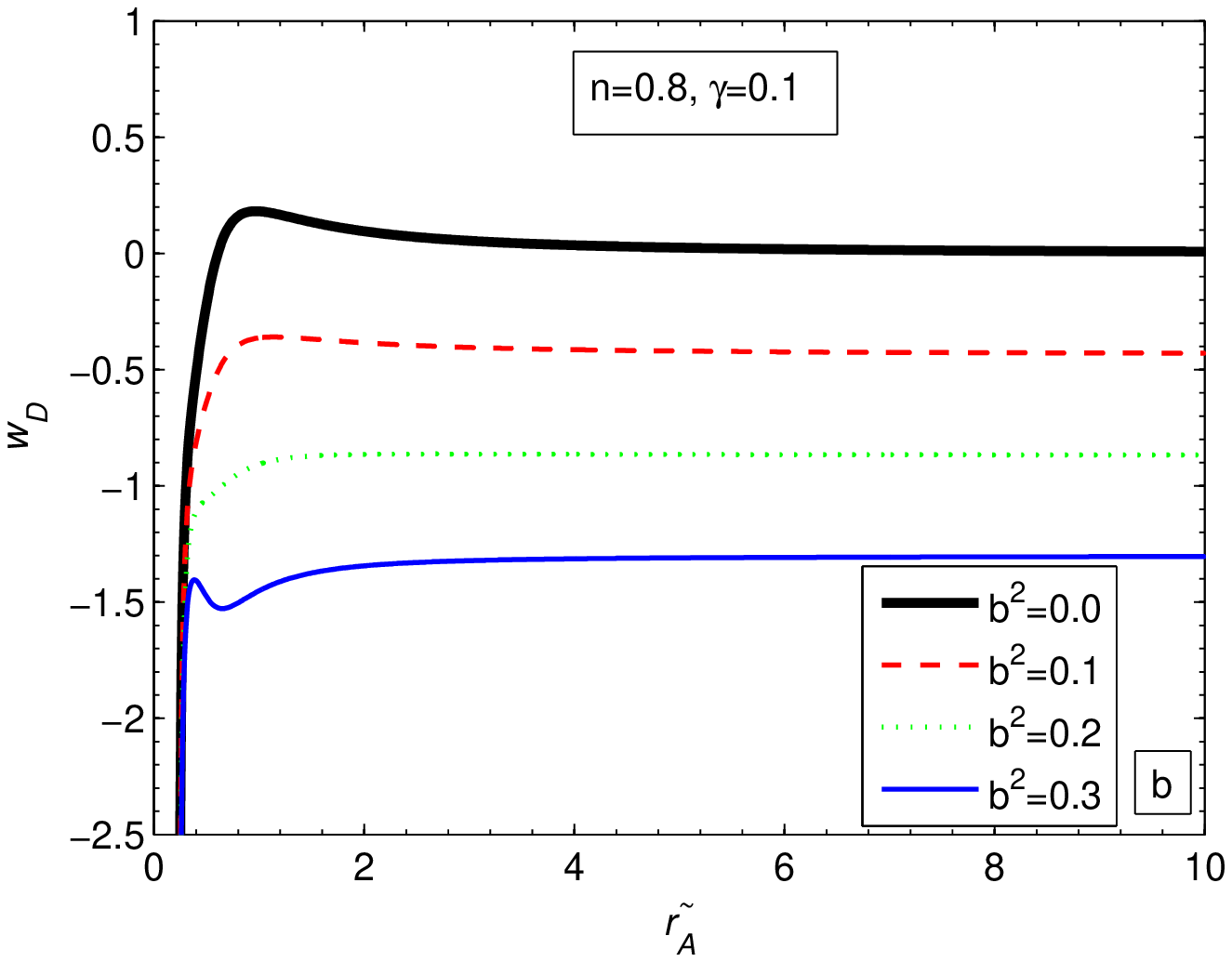}
\caption{The evolution of EoS parameter, $w_{D}$, versus of
$\widetilde{r}_{A}$ in LECHDE model, \textbf{\textit{a}:}
$0.22>\widetilde{r}_{A}>0.0$. \textbf{\textit{b}:}
$\widetilde{r}_{A}>0.23$.} \label{fig2}
\end{figure}
\end{center}

In the limiting case of ordinary HDE with $\gamma =\beta =0$, Eqs.
(\ref{eq11}, \ref{EoS}, \ref{dec2}) reduce to the following simple
forms
\begin{eqnarray}
u &=&1/n^{2}-1,  \label{eq12} \\
w_{D} &=&-(1+\frac{1}{u})\frac{\Gamma }{3H},  \label{limw} \\
q &=&-(1+\Omega _{K})-\frac{3}{2}\Omega _{D}(1+u)(\frac{\Gamma
}{3Hu}-1), \label{dec3}
\end{eqnarray}%
which have been also calculated by \cite{Sheykhi1}. In this case,
from Eq. (\ref{rp}), the radius of apparent horizon,
$\widetilde{r}_{A}$, can be obtained as
\begin{equation}
\widetilde{r}_{A}=\widetilde{r}_{A_{0}}e^{\frac{3M_{P}^2}{2}(\frac{n^2-1+b^2}{n^2-1})x}=
\widetilde{r}_{A_{0}}(1+z)^{\frac{3M_{P}^2}{2}(\frac{n^2-1+b^2}{1-n^2})}.
\label{OrA}
\end{equation}
Here we can choose $\widetilde{r}_{A_{0}}=1$ at present time: ($x=0$
or vanishing redshift, $z=0$). Therefore $\widetilde{r}_{A}$ may be
considered as a normalized horizon radius. From Eq. (\ref{OrA}), we
see that the radius of apparent horizon is increased by cosmic time
provided that $|n|>1$ or $|n|<\sqrt{1-b^2}$. From Eq. (\ref{limw}),
we see that, in the absence of interaction, we have $w_{D}=0$, but
in LECHDE model, the EoS parameter may cross the phantom divide
($w_{D}<-1$) even in the absence of interaction. In Fig. \ref{fig2},
the evolution of the EoS parameter of LECHDE in versus of
$\widetilde{r}_{A}$ is studied, both in interacting and
non-interacting modes for positive values of $\beta$, in the Planck
mass unit. We consider specially the effect of coupling constant on
behavior of $w_{D}$. As it is shown in Fig. \ref{fig2}, by choosing
the typical value of parameters of LECHDE model as:
$\gamma=0.1,~\beta=0.2,~n=0.8$, two distinct regions of
$\widetilde{r}_{A}$ are given as:

\textbf{a}: ($0.22>\widetilde{r}_{A}>0$), Fig \ref{fig2}.$a$.
Neither of interacting and non-interacting cases can drive an
expanding universe ($w_{D}>0$).

\textbf{b}: ($\widetilde{r}_{A}>0.23$), Fig \ref{fig2}.$b$. Both of
interacting and non-interacting cases may accelerate the expanding
universe and cross the phantom divide. Interacting cases always
remain under the quintessence wall, while in non-interacting mode,
the EoS parameter grows from phantom regime, $w_{D}<-1$, to positive
value of EoS parameter, ($w_{D}>-1/3$) at small values of
$\widetilde{r}_{A}<1$. Therefore the non-interacting case can not
drive the late time acceleration in our universe.

\begin{center}
\begin{figure}[]
\includegraphics[width=9cm]{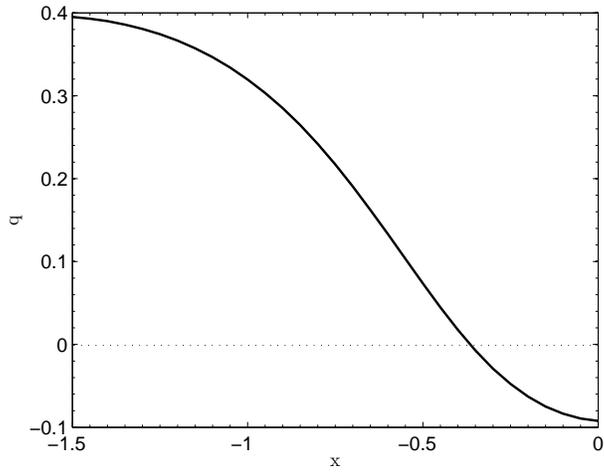}
\caption{The evolution of $q$ in versus $x=\ln{(a)}$ in LECHDE model
for ($n=0.8,~\gamma=0.1,~\beta=0.2,~b^2=.1,~\Gamma=0.04$). }
\label{fig3}
\end{figure}
\end{center}

By solving Eqs. (\ref{rp}, \ref{ed3}, \ref{dec2}, \ref{Ok})
numerically, the behavior of deceleration parameter, $q$ with
respect to $x=\ln{(a)}$, in LECHDE model, is studied. In Fig.
\ref{fig3} as we can see, the present ($x\approx0$) accelerated
stage ($q<0$ is preceded by a sufficiently long period deceleration
at the early time ($x<0$, far from $x=0$). This is compatible with
cosmic structure formation at matter dominated era and present
accelerated expansion.

The typical values of $\gamma,~\beta~,~n$ are set, so that the
function $u$ becomes positive for all studied regions and gets
$u_{0}\sim 0.4$ at present time and rich to a constant value of
order unity at the late time.
\section{Interacting ``PLECHDE" model with AH-IR-cutoff}\label{PLECHDE}

\begin{center}
\begin{figure}[]
\includegraphics[width=9cm]{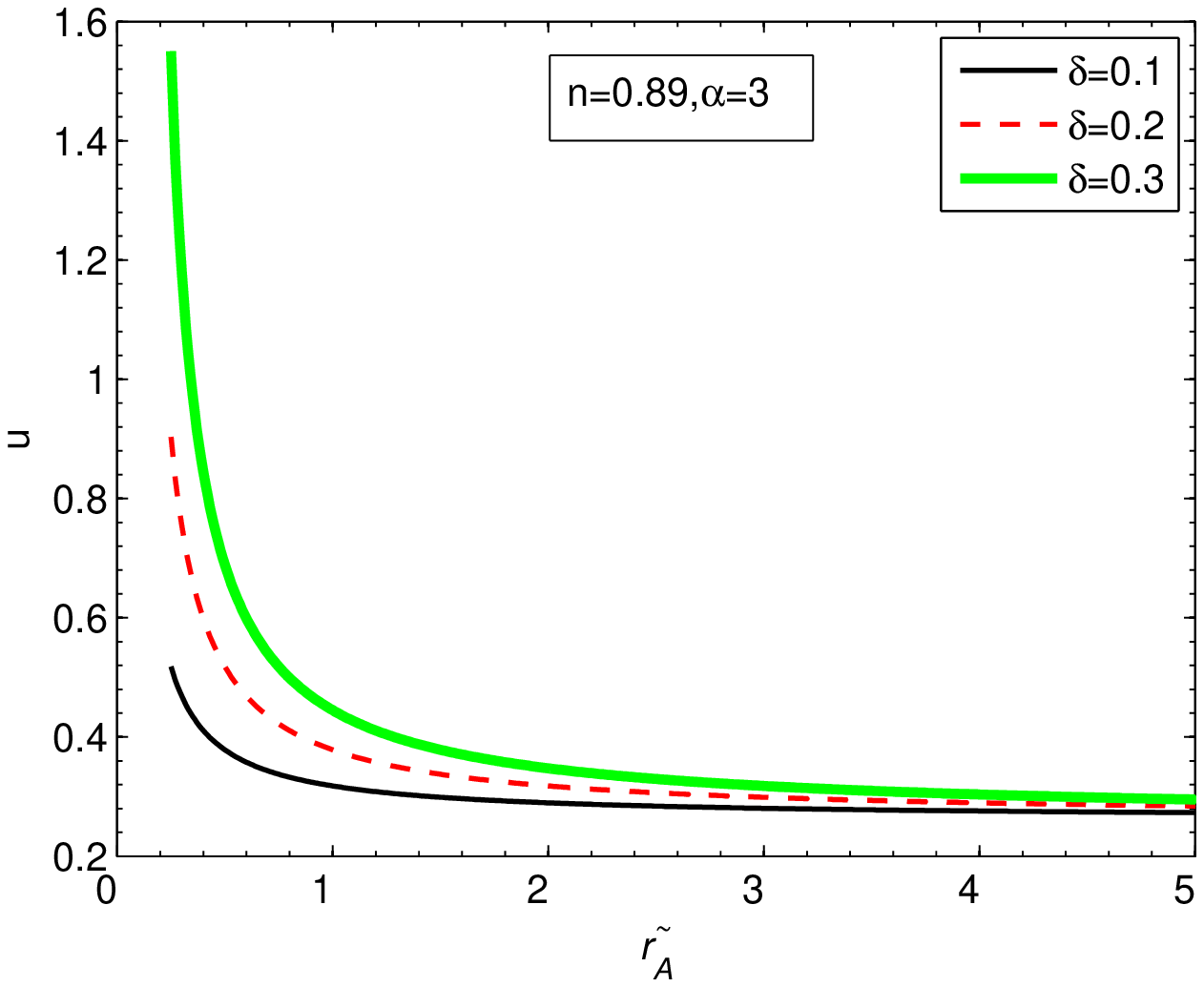}
\caption{The evolution of $u$ in versus of $\widetilde{r}_{A}$ in
PLECHDE model.} \label{fig4}
\end{figure}
\end{center}

\begin{center}
\begin{figure}[]
\includegraphics[width=9cm]{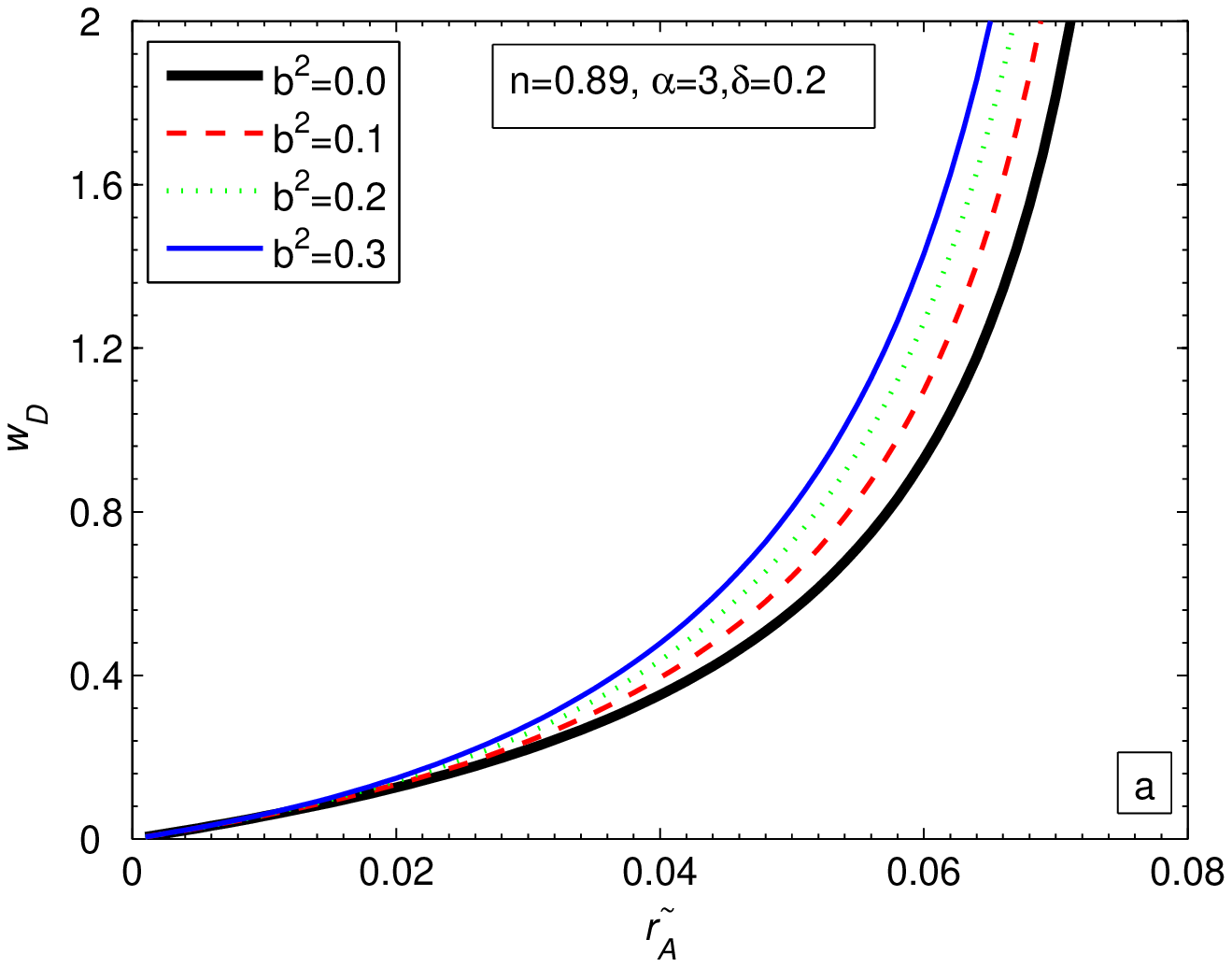}
\includegraphics[width=9cm]{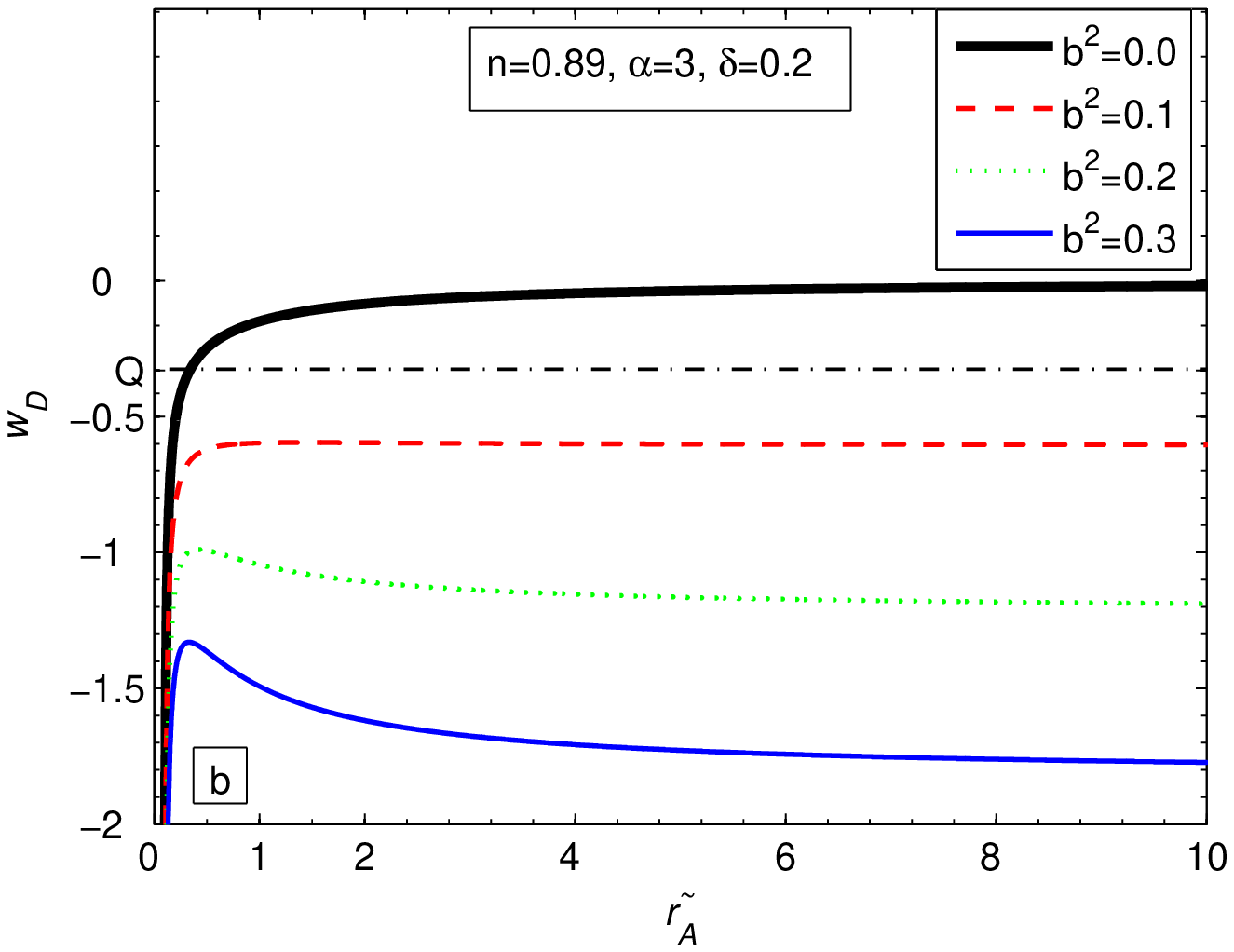}
\caption{The evolution of EoS parameter, $w_{D}$, in versus of
$\widetilde{r}_{A}$ in PLECHDE model. \textbf{\textit{a}:}
$0.08>\widetilde{r}_{A}>0.0$ and $\delta=0.2$. \textbf{\textit{b}:}
$\widetilde{r}_{A}>0.09$ and $\delta=0.2$. ``Q" the Quintessence
barrier ($w_{D}=-1/3$).} \label{fig5}
\end{figure}
\end{center}

From Eq. (\ref{endpl}), the energy density of PLECHDE with apparent
horizon, $\widetilde{r}_{A}$, as an IR-cutoff, is written as
\begin{equation}
\rho _{D}=3n^{2}M_{P}^{2}\widetilde{r}_{A}^{-2}-\delta M_{P}^{2}
\widetilde{r}_{A}^{-\alpha}, \label{EDpl}
\end{equation}
where using (\ref{rd}, \ref{EDpl}), the energy density evolution is
given by
\begin{equation}
\dot{\rho}_{D}=-3H\rho _{D}
(1+u+w_{D})\left[n^2-\frac{\alpha\delta}{6}\widetilde{r}_{A}^{2-\alpha}\right].
\end{equation}
From Eqs. (\ref{FReq}) and (\ref{EDpl}), the ratio of energy
densities, $u$, is given by
\begin{equation}
u=\frac{1}{n^2-\frac{\delta}{3} \widetilde{r}_{A}^{2-\alpha}}-1.
\label{eq21}
\end{equation}%
Also from Eqs. (\ref{EDpl}) and (\ref{eq21}), as the same as Sec.
\ref{ECHDE}, we see that at late time, for $\alpha>2$, when
$\widetilde{r}_{A}$ is large, we have $ \rho _{D}\approx
3n^{2}M_{P}^{2}\widetilde{r}_{A}^{-2}$ and the ratio of energy
densities $u$, will tend to a constant value $u\rightarrow
1/n^{2}-1$, while this is not valid for $\alpha<2$. In Fig.
\ref{fig4}, we study the behavior of $u$ in versus of
$\widetilde{r}_{A}$, for various positive values of $\delta$ and
fixed value $\alpha$. From this figure, we see that the function $u$
is descending for $\delta>0$ and the present value of $u$ is
satisfied for a typical set ($\alpha=3,~n=0.89, \delta=0.2$) at
$\widetilde{r}_{A}=1$ (present time). In this case
$u\sim\mathcal{O}(1)$, only for $\widetilde{r}_{A}> 0.3$. Also the
coincidence problem can be solved, since for some values of model
parameters, we get $u_{0}\sim\mathcal{O}(1)$, at present time, and
it finally reaches to a fixed value of order unity.
Similar to previous section, the EoS parameter $w_{D}$,
$\widetilde{r}^{\prime}_{A}$, $\Omega^{\prime}_D$ and deceleration
parameter $q$ are calculated as
\begin{eqnarray}
w_{D}&=&-\frac{1-(1+u)(n^2-\frac{\alpha\delta}{6}
\widetilde{r}_{A}^{2-\alpha}-b^2)}{1 -(n^2-\frac{\alpha\delta}{6}
\widetilde{r}_{A}^{2-\alpha})}, \label{EoSpl}\\
\widetilde{r}^{\prime}_{A}&=&\frac{3\widetilde{r}_{A}}{2}\left[\frac{1+b^2-(n^2-\frac{\delta}{3}
\widetilde{r}_{A}^{2-\alpha})}{1 -(n^2-\frac{\alpha\delta}{6}
\widetilde{r}_{A}^{2-\alpha})}\right],\label{rp2}\\
\Omega^{\prime}_D&=&-\Omega_D\left[(1+u+w_D)(3n^2-\frac{\alpha\delta}{2}\widetilde{r}_{A}^{2-\alpha}-3\Omega_D)+2\Omega_K\right],
\label{edp2}\\
 q&=&-(1+\Omega _{K})+\frac{3\Omega
_{D}}{2}\left[\frac{u-b^2(1+u)}{1 -(n^2-\frac{\alpha\delta}{6}
\widetilde{r}_{A}^{2-\alpha})}\right]. \label{decpl}
\end{eqnarray}%
The limiting case of Eqs. (\ref{eq21}, \ref{EoSpl}, \ref{decpl}),
with $\delta =0$ or large $\widetilde{r}_{A}$, has been given by
Eqs. (\ref{eq12}, \ref{limw}, \ref{dec3}). Also in this case the eq.
(\ref{rp2}) reaches to Eq. (\ref{OrA}) in the previous section. In
PLECHDE model, the EoS parameter may cross the phantom divide
($w_{D}<-1$) even in the absence of interaction. In Fig. \ref{fig5},
the EoS parameter of PLECHDE is studied both in various interacting
and non-interacting modes. As it is shown in Fig. \ref{fig5}, by
choosing the typical value of parameters of PLECHDE as:
($\alpha=3,~\delta=+0.2,~n=0.89)$, we encounter with two distinct
regions of $\widetilde{r}_{A}$ in behavior of EoS parameter as
below:

\textbf{a}: $(0.08>\widetilde{r}_{A}>0)$, Fig \ref{fig5}.$a$. We
find: ($w_{D}>0$). So the model can not drive an acceleration
expansion irrespective of interaction.

\textbf{b}: $(\widetilde{r}_{A}>0.09)$, Fig \ref{fig5}.$b$. Both of
interacting and non-interacting cases may accelerate the expansion
of the universe and the phantom divide is crossed. Interacting cases
always remains under the quintessence regime ($w_{D}<-1/3$), while
in non-interacting mode, the EoS parameter grows from phantom
regime, $w_{D}<-1$, to above the quintessence regime ($w_{D}>-1/3$)
very soon. Therefore, same as previous section, the non-interacting
case can not drive the late time acceleration.

\begin{center}
\begin{figure}[]
\includegraphics[width=9cm]{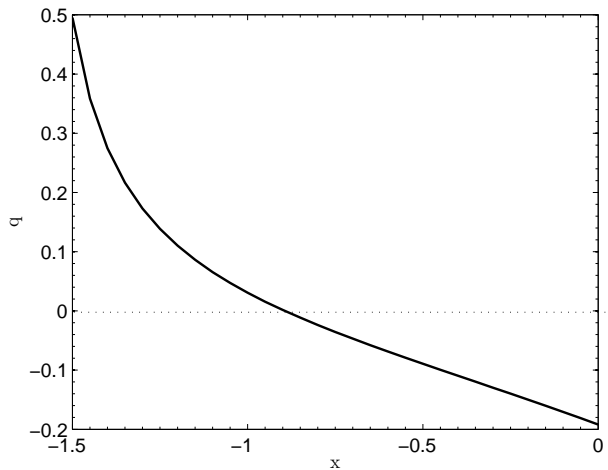}
\caption{The evolution of $q$ in versus $x=\ln{(a)}$ in PLECHDE
model for ($n=0.89,~\alpha=3,~\delta=0.2,~b^2=0.1,~\Gamma=0.04$). }
\label{fig6}
\end{figure}
\end{center}

Now we want to study the deceleration parameter of PLECHDE model. By
solving Eqs. (\ref{rp2}, \ref{edp2}, \ref{decpl}, \ref{Ok}),
numerically, the behavior of $q$ with respect to $x$ can be studied.
In Fig. \ref{fig6}, similar to previous case, the present ($x\approx
0$) acceleration has been supported by a long period deceleration
phase at past ($x<0$).

It must be mention that, similar to previous model, the typical
values of $\alpha,~\delta~,~n$ are set, so that the function $u$
become positive for all studied regions and gets $u_{0}\sim 0.4$ at
present time.
\section{Thermodynamics of non-interacting LECHDE with AH-IR-cutoff
\label{nonint}}
 In this section we want to associate a thermodynamical
description to cosmological horizons, similar to black hole physics.
In a FRW universe enclosed by an apparent horizon, one can associate
the Hawking temperature to the horizon, which is inversely
proportional to size of the apparent horizon. We know that the FRW
universe may consist several cosmic ingredients including dark
energy, dark matter, radiation and baryonic matter. However many
cosmological evident reveal that the dark energy and matter are two
dominant components in our universe. At following, we will consider
only LECHDE and CDM components in a non-flat FRW universe enclosed
by apparent horizon. In a local thermal equilibrium, where there is
not any heat flow from the apparent horizon, the temperature of the
energy content of the universe ($T$) should be equal to the
temperature which is associated with apparent horizon ($T_{h}$). In
non equilibrium case, the heat will flow outside (inside) the
apparent horizon if the temperature of cosmic fluid is hotter
(colder) than the apparent horizon, respectively. The thermal
equilibrium state can be accessed at a finite time and therefore we
can consider a unit temperature for whole spacetime (contain DE, CDM
and AH). The equilibrium entropy of the LECHDE is connected with its
energy and pressure, $p_{D}$, through the Gibbs law of
thermodynamics
\begin{equation}
TdS_{D}=dE_{D}+p_{D}dV,  \label{FL1}
\end{equation}%
where $V=(4\pi /3)r_{A}^{3}$ is the volume of whole space up to
horizon surface and $S_{D}$ is the entropy of DE component. The
equilibrium temperature $T$, can be obtained from the surface
gravity ($\kappa _{H}$) of horizon as follows \cite{Cai1}
\begin{equation}
T=\frac{\left\vert \kappa _{H}\right\vert }{2\pi }=\frac{1}{4\pi \sqrt{-h}}%
\left\vert \partial _{a}(\sqrt{-h}h^{ab}\partial _{b}\widetilde{r}%
)\right\vert .  \label{T1}
\end{equation}%
From this equation, the temperature of apparent horizon is
calculated as
\begin{equation}
T=\frac{1}{2\pi \widetilde{r}_{A}}\left( 1-\frac{\dot{\widetilde{r}}_{A}}{2H%
\widetilde{r}_{A}}\right) .  \label{T2}
\end{equation}%
Following Cai and Kim \cite{Cai1}, the apparent horizon radius
$\widetilde{r}_{A}$ should be regarded to have a fixed value in
thermal equilibrium. It means that
$\dot{\widetilde{r}}_{A}\approx0.$ Thus the temperature is given by
\begin{equation}
T=1/(2\pi \widetilde{r} _{A}^{(0)}). \label{T1}
\end{equation}
Now from Eq. (\ref{FL1}), we have
\begin{equation}
TdS=\rho _{D}(1+w_{D})dV+Vd\rho _{D},
\end{equation}%
and by using Eq. (\ref{ED}), we can obtain%
\begin{eqnarray}
\frac{dS^0_{D}}{d\widetilde{r}^0_{A}}&=&\frac{8}{3}\pi ^{2}(\widetilde{r}^0_{A})^{3}%
\Big{[} 6n^{2}M_{P}^{2}(\widetilde{r}^0_{A})^{-2}\nonumber \\
&&+2\gamma (\widetilde{r}^0%
_{A})^{-4}-\rho^0 _{D}(1-3w^0_{D})\Big{]},  \label{dS1}
\end{eqnarray}
where superscript (0) denotes that the universe is in a stable
thermodynamical equilibrium state.
\section{Thermodynamics of interacting LECHDE with
AH-IR-cutoff\label{int}} In the presence of interaction, ($Q\neq
0$), the thermal equilibrium is no further maintain due to thermal
fluctuation which has been arose from decaying of dark energy to
dark matter. The conservation equations for $\rho_{m}$ and
$\rho_{D}$, have been given by Eqs. (\ref{CE1}, \ref{CE2}). In this
case, however the Gibbs law of thermodynamics may hold only
approximately for dynamical apparent horizon, the entropy affected
under a first order logarithmic correction ($S_{D}^{(1)}$) involving
temperature $T$ and the heat capacity $C$, as bellow \cite{Das}
\begin{equation}
S_{D}^{(1)}=-\frac{1}{2}\ln (CT^{2}).  \label{SC1}
\end{equation}%
Hence, the entropy should be modified as:
$S_{D}=S_{D}^{(0)}+S_{D}^{(1)}$. The heat capacity in thermal
equilibrium has been defined as: $C=T\partial S_{D}^{(0)}/\partial
T$. Using (\ref{T1}), the heat capacity can be rewritten as:
$C=-(\widetilde{r}_{A}^{0})\partial S_{D}^{(0)}/\partial
\widetilde{r}_{A}^{0}$. Using Eq. (\ref{dS1}) in
thermal equilibrium, the corrected term $S_{D}^{(1)}$ is calculated as%
\begin{eqnarray}
S_{D}^{(1)}&=&-\frac{1}{2}\ln \left[ \rho _{D}^{0}(\widetilde{r}%
_{A}^{0})^{2}(1-3w_{D}^{0})-6n^{2}M_{P}^{2}-2\gamma (\widetilde{r}%
_{A}^{0})^{-2}\right]\nonumber \\
&& -\frac{1}{2}\ln (\frac{2}{3}).
\label{SC2}
\end{eqnarray}
similar to Eq. (\ref{dS1}) with interaction, one obtains
\begin{equation}
dS_{D}=\frac{8}{3}\pi ^{2}\widetilde{r}_{A}^{3}%
\left[ 6n^{2}M_{P}^{2}\widetilde{r}_{A}^{-2}+2\gamma \widetilde{r}%
_{A}^{-4}-\rho _{D}(1-3w_{D})\right]d\widetilde{r}_{A}, \label{dS2}
\end{equation}
where from $dS_{D}=dS_{D}^{(0)}+dS_{D}^{(1)}$, we can find
\begin{eqnarray}
1-3w_{D}&=&\Big{[} 6n^{2}M_{P}^{2}\widetilde{r}_{A}^{-2}+2\gamma \widetilde{r}%
_{A}^{-4}\nonumber \\
&&-\frac{3}{8\pi ^{2}\widetilde{r}_{A}^{3}}\left(
\frac{dS_{D}^{(0)}}{
d\widetilde{r}_{A}}+\frac{dS_{D}^{(1)}}{d\widetilde{r}_{A}}\right)
\Big{]} \rho _{D}^{-1}.  \label{EoS1}
\end{eqnarray}%
From Eqs. (\ref{dS1}, \ref{SC2}), it is obtained
\begin{eqnarray}
\frac{dS_{D}^{(0)}}{d\widetilde{r}_{A}}& =&\frac{dS_{D}^{(0)}}{d\widetilde{r}%
_{A}^{0}}\frac{d\widetilde{r}_{A}^{0}}{d\widetilde{r}_{A}}=\frac{8}{3}\pi
^{2}(\widetilde{r}_{A}^{0})^{3}\Big{[} 6n^{2}M_{P}^{2}(\widetilde{r}%
_{A}^{0})^{-2}\nonumber \\
&&+2\gamma (\widetilde{r}_{A}^{0})^{-4}-\rho
_{D}^{0}(1-3w_{D}^{0})\Big{]} \frac{d\widetilde{r}_{A}^{0}}{d\widetilde{r}%
_{A}},  \\
\frac{dS_{D}^{(1)}}{d\widetilde{r}_{A}}& =&\frac{dS_{D}^{(1)}}{d\widetilde{r}%
_{A}^{0}}\frac{d\widetilde{r}_{A}^{0}}{d\widetilde{r}_{A}}=-\frac{1}{2}
\Big{\{}2\rho _{D}^{0}(\widetilde{r}_{A}^{0})(1-3w_{D}^{0})+4\gamma
(\widetilde{r}
_{A}^{0})^{-3}\nonumber \\
&&+(\widetilde{r}_{A}^{0})^{2}\frac{d}{d\widetilde{r}_{A}^{0}}%
[\rho _{D}^{0}(1-3w_{D}^{0})]\Big{\}}/\Big{[}\rho _{D}^{0}(\widetilde{r}%
_{A}^{0})^{2}(1-3w_{D}^{0})\nonumber \\
&&-6n^{2}M_{P}^{2}-2\gamma (\widetilde{r}%
_{A}^{0})^{-2}\Big{]}\frac{d\widetilde{r}_{A}^{0}}{d\widetilde{r}_{A}},
\label{Eq2}
\end{eqnarray}%
where from (\ref{EoS}) and (\ref{eq11}), we have
\begin{eqnarray}
&&1-3w_{D}=  \label{eq22} \\
&&4+3\frac{u(2\rho _{D}-3n^{2}M_{P}^{2}\widetilde{r}%
_{A}^{-2}-\gamma \widetilde{r}_{A}^{-4})-\frac{\Gamma }{3H}(1+u)\rho _{D}}{%
(1-u)\rho _{D}-3n^{2}M_{P}^{2}\widetilde{r}_{A}^{-2}-\gamma \widetilde{r}%
_{A}^{-4}},\nonumber \\
&&1-3w_{D}^{0}= \\
&&4+3u^{0}\frac{2\rho _{D}^{0}-3n^{2}M_{P}^{2}(\widetilde{r}%
_{A}^{0})^{-2}-\gamma (\widetilde{r}_{A}^{0})^{-4}}{(1-u^{0})\rho
_{D}^{0}-3n^{2}M_{P}^{2}(\widetilde{r}_{A}^{0})^{-2}-\gamma (\widetilde{r}%
_{A}^{0})^{-4}},\nonumber \\
&&\frac{du^{0}}{d\widetilde{r}_{A}^{0}} =-(1+u^{0})\left[ \frac{2}{%
\widetilde{r}_{A}^{0}}+\frac{d}{d\widetilde{r}_{A}^{0}}\ln (\rho _{D}^{0})%
\right].
\end{eqnarray}%
Now, we want to find a relation between the interaction term and the
thermal fluctuation. For this purpose, by comparing Eqs.(\ref{EoS1},
\ref{eq22}), the interaction term can be calculated with respect to
thermal fluctuation as
\begin{eqnarray}
&&\frac{\Gamma }{3H}=\frac{2}{3(1+u)\rho _{D}^{2}}{\Big \{}(2\rho
_{D}-3n^{2}M_{P}^{2}\widetilde{r}_{A}^{-2}-\gamma \widetilde{r}
_{A}^{-4})\label{CC} \\
&&\left( (1+\frac{u}{2})\rho _{D}-3n^{2}M_{P}^{2}\widetilde{r}
_{A}^{-2}-\gamma \widetilde{r}_{A}^{-4}\right)+(\frac{d\widetilde{r}_{A}}{d\widetilde{r}_{A}^{0}})\times  \nonumber \\
&&\frac{3\widetilde{r} _{A}^{0}}{32\pi
^{2}\widetilde{r}_{A}^{3}}\frac{[(1-u)\rho
_{D}-3n^{2}M_{P}^{2}\widetilde{r}_{A}^{-2}-\gamma
\widetilde{r}_{A}^{-4}]}{ 6n^{2}M_{P}^{2}+2\gamma
(\widetilde{r}_{A}^{0})^{-2}-\rho _{D}^{0}(
\widetilde{r}_{A}^{0})^{2}(1-3w_{D}^{0})}\times  \nonumber \\
&&{\Big [}\frac{16}{3}\pi ^{2}\left( 6n^{2}M_{P}^{2}+2\gamma
(\widetilde{r} _{A}^{0})^{-2}-\rho
_{D}^{0}(\widetilde{r}_{A}^{0})^{2}(1-3w_{D}^{0})\right)
^{2}  \nonumber \\
&&+4\gamma (\widetilde{r}_{A}^{0})^{-4}+2\rho
_{D}^{0}(1-3w_{D}^{0})+\widetilde{r}_{A}^{0}\frac{d}{d\widetilde{r}
_{A}^{0}}[\rho _{D}^{0}(1-3w_{D}^{0}){\Big ]}{\Big \}}.\nonumber
\end{eqnarray}%
In limiting case, for ordinary HDE ($\gamma =\beta =0)$, where
$w_{D}^{0} =0$ and $\rho
_{D}=3n^{2}M_{P}^{2}\widetilde{r}_{A}^{-2}$, from Eqs. (\ref {limw},
\ref{CC}), we can obtain
\begin{equation}
\frac{\Gamma }{3H} =\frac{1-n^{2}}{3}\left[ 1-\widetilde{r}_{A}^{0}\frac{d%
}{d\widetilde{r}_{A}^{0}}\ln (\widetilde{r}_{A})\right] .
\end{equation}

\section{Conclusion}
In this paper the logarithmic and power-law entropy-corrected
version of interacting HDE with AH-IR-cutoff in a non-flat universe
enclosed by apparent horizon have been studied. In fact we
generalized the ordinary HDE model by considering the entropy
correction due to fluctuation of spacetime and AH-IR-cutoff. In
LECHDE model, corrections are restricted to the leading order
correction which contains the logarithmic of area. In PLECHDE model,
the correction is based on the gravitational fluctuations which
affect the area law of entropy to a fractional power of area, which
is arisen by entanglement of quantum field theory. The ratio of dark
matter to dark energy densities $u$, EoS parameter $w_D$ and
deceleration parameter $q$ have been obtained. We showed that the
cosmic coincidence is satisfied for appropriate model parameters. In
dealing with cosmic coincidence problem, we found an appropriate set
of values for LECHDE model as: ($\gamma=0.1,~\beta=0.2,~n=0.8$) and
for PLECHDE model as: ($n=0.89~\alpha=3,~\delta=0.2$). These
parameters have been chosen in order to get $u_{0}\sim 0.4$ and
finally, reaches slowly to a constant value of order unity.  By
studying the effect of interaction in EoS parameter, we saw that the
phantom divide may be crossed and also find that the interacting
models can drive an acceleration expansion at the present and
future, while in non-interacting case, this expansion can happen
only at the early time. The graphs of deceleration parameter for
interacting models, showed that the present acceleration expansion
is preceded by a sufficiently long period deceleration at past.

Moreover, the thermodynamical interpretation of interaction between
LECHDE and dark matter was described. Based on the Gibbs law of
thermodynamics, for dark energy sector of the universe in
non-interacting case, we calculated a differentiation of entropy of
DE with respect to $\widetilde{r}_{A}$. Although in the absence of
interaction between dark energy and dark matter, these two dark
components conserved separately, while by imposing an interaction
term, a stable fluctuation around equilibrium is expectable.
Therefore, in the interacting case, where the entropy affected under
a first order logarithmic correction, we obtained a relation between
the interaction term and thermal fluctuation in a non-flat universe
enclosed by the apparent horizon. Also in limiting case for ordinary
HDE, the relation of interaction term versus thermal fluctuation was
calculated.

\acknowledgments {We are grateful to the referee for valuable
comments and suggestions, which have allowed us to improve this
paper significantly. we sincerely thank Prof. Ahmad Sheykhi for
constructive comments on an earlier draft of this paper.}

\end{document}